\documentclass[nohyperref]{article}

\usepackage{microtype}
\usepackage{graphicx}
\usepackage{subfigure}
\usepackage{booktabs} 

\usepackage{hyperref}

\usepackage[accepted]{icml2023}

\usepackage{amsmath}
\usepackage{amssymb}
\usepackage{mathtools}
\usepackage{amsthm}
\usepackage{ulem}

\usepackage[capitalize,noabbrev]{cleveref}

\theoremstyle{plain}

\theoremstyle{definition}

\theoremstyle{remark}

\usepackage[textsize=tiny]{todonotes}

\icmltitlerunning{A Conditional Normalizing Flow for Accelerated Multi-Coil MR Imaging}

 \renewcommand{\eqref}[1]{(\ref{eq:#1})}
 \newcommand{\secref}[1]{Sec.\ \ref{sec:#1}}
 \newcommand{\tabref}[1]{Table~\ref{tab:#1}}

 \newcommand{\textr}[1]{\textcolor{black}{#1}}
 \newcommand{\textb}[1]{\textcolor{black}{#1}}

 \renewcommand{\hat}{\widehat}
 \renewcommand{\bar}{\overline}
 \newcommand{\defn}{\triangleq}

 \newcommand{\tvec}[1]{\ensuremath{\Tilde{\boldsymbol{#1}}}}

 \newcommand{\hvec}[1]{\ensuremath{\Hat{\boldsymbol{#1}}}}
 \renewcommand{\vec}[1]{\ensuremath{\boldsymbol{#1}}}

 \newcommand{\Real}{{\mathbb{R}}}
 \newcommand{\Complex}{{\mathbb{C}}}

 \newcommand{\of}[1]{^{\scriptscriptstyle (#1)}}

 \newcommand{\tran}{^{\top}}
 \newcommand{\herm}{^\textsf{H}}

 \DeclareMathOperator{\E}{E}

 \DeclareMathOperator{\blkdiag}{blkdiag}

 \DeclareMathOperator{\nullsp}{null}

 \newcommand{\true}{_\mathsf{true}}
 
 \newcommand{\map}{_\mathsf{MAP}}

\begin{document}
\setlength{\arraycolsep}{0.5mm}
\newcommand{\spc}{-3mm} 

\twocolumn[
\icmltitle{A Conditional Normalizing Flow for Accelerated Multi-Coil MR Imaging}



\icmlsetsymbol{equal}{*}

\begin{icmlauthorlist}
\icmlauthor{Jeffrey Wen}{osu_ece}
\icmlauthor{Rizwan Ahmad}{osu_bme}
\icmlauthor{Philip Schniter}{osu_ece}

\end{icmlauthorlist}

\icmlaffiliation{osu_ece}{Dept. of ECE, The Ohio State University, Columbus, OH 43210, USA.}
\icmlaffiliation{osu_bme}{Dept. of BME, The Ohio State University, Columbus, OH 43210, USA}

\icmlcorrespondingauthor{Jeffrey Wen}{wen.254@osu.edu}

\icmlkeywords{Machine Learning, MRI, Normalizing Flow, Inverse Problems}

\vskip 0.3in
]



\printAffiliationsAndNotice{}  

\begin{abstract}
Accelerated magnetic resonance (MR) imaging attempts to reduce acquisition time by collecting data below the Nyquist rate. 
As an ill-posed inverse problem, many plausible solutions exist, yet the majority of deep learning approaches generate only a single solution. 
We instead focus on sampling from the posterior distribution, which provides more comprehensive information for downstream inference tasks. 
To do this, we design a novel conditional normalizing flow (CNF) that \textb{infers the signal component in the measurement operator's nullspace, which is later combined with measured data to form complete images}. 
Using fastMRI brain and knee data, we demonstrate fast inference and accuracy that surpasses recent posterior sampling techniques for MRI.
Code is available at \url{https://github.com/jwen307/mri_cnf}

\end{abstract}

\section{Introduction}
Magnetic resonance imaging (MRI) is a routine diagnostic imaging tool that has the potential to provide high-quality soft-tissue images without exposure to ionizing radiation. 
However, MRI exams are generally time-consuming, which reduces throughput, compromises patient comfort, and increases the likelihood of artifacts from patient motion. 
Scan time can be reduced by sampling below the Nyquist rate, but this makes the image \textr{reconstruction} process more challenging. 
Hence, recovering high-accuracy images from highly subsampled MRI scans has become an active area of research \cite{Knoll:SPM:20}. 

Many approaches have been proposed to recover MR images from subsampled measurements.  
Parallel imaging, which is available on all commercial scanners, takes advantage of the availability of multiple receiver coils. 
After estimating coil-sensitivity maps or interpolation kernels, methods like SENSE \cite{Prussmann:MRM:99} and GRAPPA \cite{Griswold:MRM:02} can use subsampled data from multiple coils to remove aliasing artifacts in the final reconstruction. 
However, parallel imaging alone can typically allow only two- to three-fold acceleration of the acquisition process. 
For higher acceleration, methods based on compressed-sensing (CS) have been proposed \cite{Lustig:MRM:07}. 
The CS methods are framed as iteratively minimizing the sum of a data-fidelity term and a regularization term, where the regularization term incorporates prior knowledge about the images. 
The prior knowledge could be that the true images are sparse in some transform domain, as in traditional CS, or that the true images are preserved by some denoising function, as in ``plug-and-play'' recovery \cite{Ahmad:SPM:20}.
Deep neural networks have also been proposed for MR image recovery, based on end-to-end approaches like \cite{Zbontar:18, Eo:MRM:18, Sriram:MICCAI:20} or algorithmic unrolling \cite{Hammernik:MRM:18}.
Yet another approach, known as compressed sensing with a generative model (CSGM) \cite{Bora:ICML:17}, trains a deep image generator and then optimizes its input to give the image that, after application of the forward model, best matches the measurements. 

Although they achieve high reconstruction quality, the aforementioned methods provide only a point estimate. 
Yet, accelerated MRI is an ill-posed inverse problem, where there exist many possible reconstructions that are consistent with a given prior and set of subsampled measurements. 
Since small variations in image content can impact the final diagnosis, it is crucial for radiologists to know whether a visual structure is truly reflective of the patient anatomy or merely an imaging artifact.
Problems of this form fall into the realm of uncertainty quantification (UQ) \cite{Abdar:IF:21}.

One approach that facilitates UQ is Bayesian imaging, where the goal is not to compute a single ``good'' image estimate but rather to sample from the posterior distribution.
The availability of a large batch of posterior samples enables many forms of UQ.
For example, a simple approach is to generate the pixel-wise standard-deviation map, which quantifies which pixels are more trustworthy. 
A more involved approach is to construct a hypothesis test for the absence of a particular (multi-pixel) visual structure \cite{Repetti:JIS:19}.
\textb{In this paper, we focus on the task of sampling from the posterior, which facilitates future work that uses those samples for uncertainty quantification, adaptive sampling \cite{Sanchez:NIPSW:20}, counterfactual diagnosis \cite{Chang:ICLR:19}, or other applications.}

There exist several deep-learning based approaches to sample from the posterior, including those based on 
conditional generative adversarial networks (CGANs) \cite{Isola:CVPR:17,Adler:18}, 
conditional variational autoencoders (CVAEs) \cite{Edupuganti:TMI:20,Tonolini:JMLR:20}, 
conditional normalizing flows (CNFs) \cite{Ardizzone:19,Winkler:19},
and score/Langevin/diffusion-based approaches \cite{Kadkhodaie:20,Laumont:JIS:22,Ho:NIPS:20}. 
In this paper, we focus on the CNF approach.
Compared to the other methods, CNFs yield rapid inference and require only simple, likelihood-based training. 
In a recent super-resolution (SR) contest \cite{Lugmayr:CVPR:22}, a CNF (by Song et al.\ \yrcite{Song:CVPRW:22}) won, beating all CGAN, CVAE, and diffusion-based competitors.

Inspired by the success of CNFs in SR, 
we design 
the first CNF for accelerated multi-coil MRI.
Previous applications of CNFs to MRI \cite{Denker:JI:21} showed competitive results but were restricted to single-coil recovery of magnitude images. 
\textb{As the vast majority of modern MRI scanners capture multi-coil data, the extension to multi-coil, complex-valued data is crucial for real-world adoption. However, the order-of-magnitude increase in dimensionality makes this transition non-trivial.}
For this purpose, we propose a novel CNF that infers only the signal component in the nullspace of the measurement operator and combines its output with the measured data to generate complete images. 
Using fastMRI brain and knee data, we demonstrate that our approach outperforms existing posterior samplers based on CGANs \cite{Adler:18} and MRI-specific score/Langevin-based approaches \cite{Jalal:NIPS:21,Chung:MIA:22} in almost all accuracy metrics, while retaining fast inference and requiring minimal hyperparameter tuning.

\section{Background}

\subsection{Measurement Model}

In MRI, measurements of the $D$-pixel true image $\vec{i}\true\in\Complex^D$ are collected in the spatial Fourier domain, known as the ``k-space.''
In a multi-coil system with $C$ coils, measurements from the $c$th coil can be written as
\begin{equation}
\vec{k}_{c} = \vec{P F S}_c \vec{i}\true + \vec{\epsilon}_{c} \in \Complex^M
\label{eq:kc} ,
\end{equation}
where 
$\vec{P} \in \Real^{M \times D}$ is a sampling matrix containing $M$ rows of the $D \times D$ identity matrix $\vec{I}$, 
$\vec{F}$ is the $D \times D$ 2D unitary discrete Fourier transform (DFT) matrix,
$\vec{S}_c \in \Complex^{D \times D}$ is the coil-sensitivity map of the $c$th coil,
and
$\vec{\epsilon}_c \in \Complex^M$ is measurement noise.
We will assume that \textr{$\{\vec{S}_c\}_{c=1}^C$} have been obtained from ESPIRiT \cite{Uecker:MRM:14}, in which case $\sum_{c=1}^C \vec{S}_c\herm\vec{S}_c=\vec{I}$.
In the case of single-coil MRI, $C=1$ and $\vec{S}_1=\vec{I}$.

We now rewrite the model in terms of the ``coil images'' $\vec{x}_c \defn \vec{S}_c\vec{i}$ and their corresponding ``zero-filled'' estimates $\vec{y}_c \defn \vec{F}\herm\vec{P}\tran\vec{k}_c$, and then stack all the coils together via $\vec{x}\true \defn [\vec{x}_1\tran,\dots,\vec{x}_C\tran]\tran$ and $\vec{y} \defn [\vec{y}_1\tran,\dots,\vec{y}_C\tran]\tran$ to obtain 
\begin{align}
\vec{y} 
&= \vec{Ax}\true + \vec{\varepsilon} 
\label{eq:y} ,
\end{align}
with \textr{$\vec{\varepsilon} = [(\vec{F}\herm\vec{P}\tran \vec{\epsilon}_1)\tran,\dots,(\vec{F}\herm\vec{P}\tran \vec{\epsilon}_C)\tran]\tran$} and forward operator
\begin{align}
\vec{A}
&= \blkdiag\big\{\vec{F}\herm\vec{P}\tran\vec{PF}, \dots, \vec{F}\herm\vec{P}\tran\vec{PF} \big\}
\label{eq:A} .
\end{align}
To perform image recovery, one can first compute $\vec{y}$, then estimate $\hvec{x}$ from $\vec{y}$, and finally either ``coil-combine'' to yield a complex-valued image estimate 
\begin{align}
\hvec{i} = [\vec{S}_1\herm,\dots,\vec{S}_C\herm]\hvec{x}
\label{eq:sense}
\end{align}
or perform root-sum-of-squares (RSS) reconstruction to obtain a magnitude-only image estimate 
\begin{align}
\textr{|\hvec{i}| = \textstyle \sqrt{\sum_{c=1}^C |\hvec{x}_c|^2}}
\label{eq:rss}.
\end{align}
In the ``fully sampled'' case, $M=D$ and so $\vec{y}=\vec{x}\true+\vec{\varepsilon}$.
But fully sampled acquisition is very slow, and so we are interested in accelerated MRI, where one collects $M \textr{<} D$ measurements per coil to save time.
This gives an ``acceleration factor'' of $R\defn D/M$, but it makes $\vec{A}$ rank deficient.
In this latter case, accurate recovery of $\vec{x}\true$ requires the use of prior information about $\vec{x}\true$, such as the knowledge that $\vec{x}\true$ is a vector of MRI coil images.


\subsection{Posterior Sampling}


In the case of MRI, the posterior distribution that we would ultimately like to sample from is $p_{i|k}(\cdot|\vec{k})$, where $\vec{k}\defn[\vec{k}_1\tran,\dots,\vec{k}_C\tran]$.
Equivalently, we could consider $p_{i|y}(\cdot|\vec{y})$ since $\vec{y}$ and $\vec{k}$ contain the same information.
Another option is to sample from {$p_{x|y}(\cdot|\vec{y})$ and then use \eqref{sense} or \eqref{rss} to \textr{combine coil images into a single image.}
We take the latter approach.

For CNFs and CGANs, posterior sampling is accomplished by designing a neural network that maps samples from an easy-to-generate latent distribution (e.g., white Gaussian) to the target distribution (i.e., the distribution of $\vec{x}$ given $\vec{y}$, with density $p_{x|y}$).
Once that network is trained, sample generation is extremely fast.
For Langevin dynamics, an algorithm is run for hundreds or thousands of iterations to generate each sample, and each iteration involves calling a neural network.  
Consequently, the inference time is much \textr{longer} than that of CNFs and CGANs.

\subsection{Conditional Normalizing Flows} \label{sec:CNF}
Normalizing flows (NF) \cite{Dinh:ICLRW:15, Dinh:ICLR:17, Kingma:NIPS:18, Papamakarios:JMLR:21} have emerged as powerful generative models capable of modeling complex data distributions. 
Normalizing flows learn an invertible mapping between a target data distribution and a simple latent distribution, generally a Gaussian. 
More concretely, for a latent sample $\vec{z}$ drawn from the latent distribution $p_{z}$, the normalizing flow defines an invertible transformation $f_{\vec{\theta}}(\cdot): \Real^{Q} \rightarrow \Real^{Q}$. 
This transformation is parameterized by $\vec{\theta}$, and $\vec{x} = f_{\vec{\theta}}(\vec{z})$ defines a sample in the target data domain. 
This mapping of the latent distribution induces a probability in the target data domain with a probability density derived from the change-of-variable formula
\begin{equation}
\hat{p}_x(\vec{x};\vec{\theta}) = p_{z}(f_{\vec{\theta}}^{-1}(\vec{x})) \bigg| \det \bigg( \frac{\partial f_{\vec{\theta}}^{-1}(\vec{x})}{\partial \vec{x}} \bigg) \bigg| ,
\end{equation}
where $\det(\cdot)$ denotes the determinant. 
The goal of the normalizing flow is to approximate the underlying data distribution $p_{x}$ with $\hat{p}_x(\cdot;\vec{\theta})$. 
Given a set of data samples $\{ \vec{x}\of{i} \}_{i=1}^{N}$, the parameters $\vec{\theta}$ can be fit using a maximum likelihood loss
\begin{align}
&L(\vec{\theta}) 
= \sum_{i=1}^{N} \ln \hat{p}_x(\vec{x}\of{i};\vec{\theta}) \\
&= \sum_{i=1}^{N} \ln p_{z}(f_{\vec{\theta}}^{-1}(\vec{x}\of{i})) + \ln\!\bigg| \det\!\bigg(\! \frac{\partial f_{\vec{\theta}}^{-1}(\vec{x}\of{i})}{\partial \textr{\vec{x}^{(i)}}} \!\bigg) \bigg|
\end{align}
Once the training is complete, samples from the target distribution can be rapidly generated by drawing samples from the latent distribution and passing them through the normalizing flow $f_{\vec{\theta}}$. 

It is worth noting that \textr{maximizing} $L(\vec{\theta})$ is equivalent to minimizing the Kullback-Leibler (KL) divergence between $\hat{p}_x(\cdot;\vec{\theta})$ and $p_{x}$ \cite{Papamakarios:JMLR:21}, which aligns with the goal of approximating $p_{x}$ with $\hat{p}_x(\cdot;\vec{\theta})$. 
The maximum-likelihood loss provides stable training with minimal hyperparameter tuning and has been shown to be robust to mode collapse.

Conditional normalizing flows (CNFs) \cite{Ardizzone:21} generalize normalizing flows by adding a conditioning signal $\vec{y}$. 
With the CNF denoted as $h_{\vec{\theta}}(\cdot,\cdot): \Real^{Q}\times \Real^Q \rightarrow \Real^{Q}$, the forward process from the latent domain to the data domain is given by $\vec{x} = h_{\vec{\theta}}(\vec{z},\vec{y})$. 
For complex-valued, multi-coil MRI, we have $Q=2CD$.
The inclusion of $\vec{y}$ alters the objective of the CNF to approximating the unknown posterior distribution $p_{x|y}(\cdot|\vec{y})$ with $\hat{p}_{x|y}(\cdot|\vec{y};\vec{\theta})$. 
As before, the change-of-variable formula implies the induced distribution
\begin{equation}
    \hat{p}_{x|y}(\vec{x}|\vec{y};\vec{\theta}) = p_{z}(h_{\vec{\theta}}^{-1}(\vec{x},\vec{y})) \bigg|\!\det\!\bigg( \frac{\partial h_{\vec{\theta}}^{-1}(\vec{x},\vec{y})}{\partial \vec{x}} \bigg) \bigg| 
\label{eq:px|y},
\end{equation}
where $h_{\vec{\theta}}^{-1}$ refers to the inverse mapping of $h_{\vec{\theta}}$ with respect to its first argument. 

Given a dataset $\{(\vec{x}\of{i}, \vec{y}\of{i})\}_{i=1}^{N} $, the maximum likelihood loss can be utilized to optimize the parameters $\vec{\theta}$
\begin{align}
&L(\vec{\theta}) 
= \sum_{i=1}^{N} \ln \hat{p}_{x|y}(\vec{x}\of{i} | \vec{y}\of{i};\vec{\theta}) \\
&\!= \sum_{i=1}^{N} \ln p_{z}(h_{\vec{\theta}}^{-1}(\vec{x}\of{i},\vec{y}\of{i}))
\!+\! \ln \!\bigg| \!\det \!\bigg( \frac{\partial h_{\vec{\theta}}^{-1}(\vec{x}\of{i},\vec{y}\of{i})}{\partial \vec{x}\of{i}} \bigg) \bigg|
\label{eq:conditional loss}
\end{align}
CNFs have shown promising performance in solving inverse problems, such as super-resolution \cite{Lugmayr:ECCV:20, Kim:21, Song:CVPRW:22}, making it an exciting avenue of exploration for accelerated MRI. 
Denker et al.\ \yrcite{Denker:JI:21} developed a CNF for single-coil, magnitude-only knee images. 
This study showed promising initial results, but the limited scope did not demonstrate performance in the more realistic multi-coil, complex-valued domain. 
\textb{As this transition increases the dimensionality by an order of magnitude, non-trivial architectural changes are required.}
In this paper, we build on the latest advances in CNFs to create a method that is capable of generating high-quality posterior samples of multi-coil, complex-valued MRI images.

\begin{figure*}[t]
\centerline{\includegraphics[width=\textwidth,trim=10 10 8 7,clip]{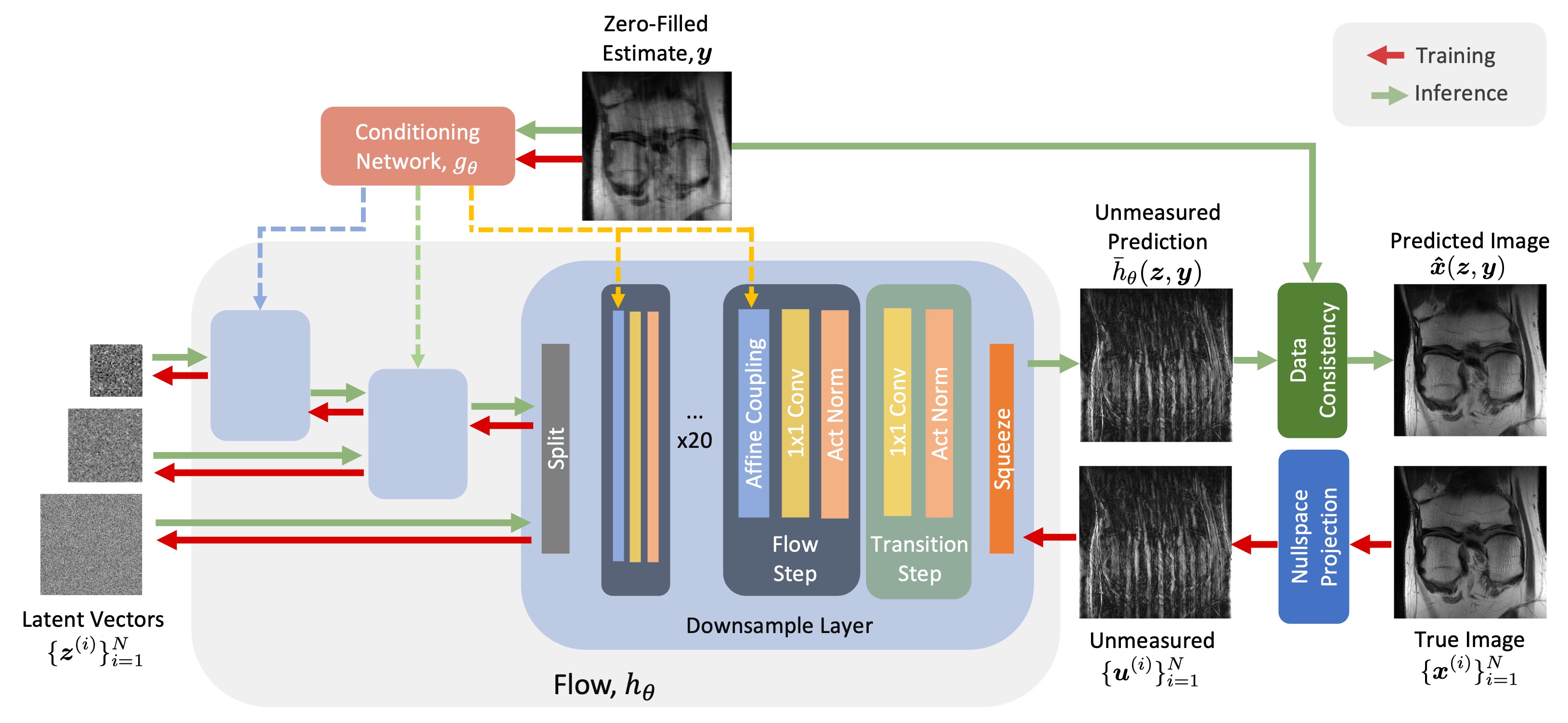}}
\vspace{\spc}
\caption{The architecture of our CNF. The conditioning network $g_{\vec{\theta}}$ takes in multi-coil zero-filled image estimates $\vec{y}$ and outputs features used by the flow model $h_{\vec{\theta}}$. The flow learns an invertible mapping between Gaussian random samples $\vec{z}\of{i}$ and images $\vec{u}\of{i}$ that are the projections of the training images $\vec{x}\of{i}$ onto the non-measured subspace. 
}
\label{fig:architecture}
\end{figure*}

\section{Method} \label{sec:Method}

Our CNF consists of two networks, a conditioning network $g_{\vec{\theta}}$ and a conditional flow model $h_{\vec{\theta}}$. 
The conditioning network takes the vector of zero-filled (ZF) coil-images $\vec{y}$ as input and produces features that are used as conditioning information by the flow model $h_{\vec{\theta}}$. 
Aided by the conditioning information, $h_{\vec{\theta}}$ learns an invertible mapping between samples in the latent space and those in the image space. 
Using the notation of Sec.\ \ref{sec:CNF}, our overall CNF takes the form 
\begin{align}
\bar{h}_{\vec{\theta}}(\vec{z},\vec{y}) \defn h_{\vec{\theta}}(\vec{z},g_{\vec{\theta}}(\vec{y})).
\end{align}

Recently, advancements of CNFs in the super-resolution literature have revealed useful insights for more general inverse problems. 
First, Lugmayr et al.\ \yrcite{Lugmayr:ECCV:20} suggested the use of a pretrained, state-of-the-art point-estimate network for the conditioning network $g_{\vec{\theta}}$. 
This network is then trained jointly with $h_{\vec{\theta}}$ using the loss in \eqref{conditional loss}. 
This approach provides a functional initialization of $g_{\vec{\theta}}$ and allows $g_{\vec{\theta}}$ to learn to provide features that are useful for the maximum-likelihood training objective. 
We utilize a UNet from \cite{Zbontar:18} for $g_{\vec{\theta}}$ since it has been shown to perform well in accelerated MRI. 
We first pre-train $g_{\vec{\theta}}$ for MRI recovery, and later we jointly train $g_{\vec{\theta}}$ and $h_{\vec{\theta}}$ together. 

Song et al.\ \yrcite{Song:CVPRW:22} demonstrated the benefits of using ``frequency-separation" when training a CNF for super-resolution. 
The authors argue that the low-resolution conditional image already contains sufficient information about the low-frequency components of the image, so the CNF can focus on recovering only the high-frequency information. 
The CNF output is then added to an upsampled version of the conditional image to yield an estimate of the full image. 

We now generalize the frequency-separation idea to arbitrary linear models of the form $\vec{y}=\vec{Ax}\true+\textr{\vec{\varepsilon}}$ from \eqref{y} and apply the resulting procedure to MRI. 
Notice that \eqref{y} implies 
\begin{align}
\vec{A}^+\vec{y} = \vec{A}^+\vec{Ax}\true + \vec{A}^+\textr{\vec{\varepsilon}}
\end{align}
where $(\cdot)^+$ denotes the pseudo-inverse.
Here, $\vec{A}^+\vec{Ax}\true$ is recognized as the projection of $\vec{x}\true$ onto the row-space of $\vec{A}$, which we will refer to as the ``measured space.''
Then 
\begin{align}
\vec{u}\true \defn (\vec{I}-\vec{A}^+\vec{A})\vec{x}\true
\end{align}
would be the projection of $\vec{x}\true$ onto its orthogonal complement, which \textb{we refer to as the ``nullspace.''}
Assuming that the \textb{nullspace} has dimension $>0$, we propose to construct an estimate $\hvec{x}$ of $\vec{x}\true$ with the form
\begin{align}
\hvec{x}(\vec{z},\vec{y})
= (\vec{I}-\vec{A}^+\vec{A})\bar{h}_{\vec{\theta}}(\vec{z},\vec{y}) + \vec{A}^+\vec{y} 
\label{eq:xhat},
\end{align}
where $\bar{h}_{\vec{\theta}}(\vec{z},\vec{y})$ is our CNF-generated estimate of $\vec{u}\true$ and the $(\vec{I}-\vec{A}^+\vec{A})$ in \eqref{xhat} strips off any part of $\bar{h}_{\vec{\theta}}(\vec{z},\vec{y})$ that has leaked into the measured space.
\textb{A similar approach was used in \cite{Sonderby:ICLR:17} for point estimation.} 
Given training data $\{(\vec{x}\of{i},\vec{y}\of{i})\}_{i=1}^N$, the CNF \textr{$\bar{h}_{\vec{\theta}}(\cdot,\cdot)$} is trained to map code vectors $\vec{z}\of{i}\sim p_z$ to \textb{the nullspace projections}
\begin{align}
\vec{u}\of{i}\defn (\vec{I}-\vec{A}^+\vec{A})\vec{x}\of{i}
\label{eq:ui}
\end{align} 
using the measured data $\vec{y}\of{i}$ as the conditional information. 
As a result of \eqref{xhat}, the reconstructions $\hvec{x}$ agree with the measurements $\vec{y}$ in that $\vec{A}\hvec{x}=\vec{y}$.
However, this also means that $\hvec{x}$ inherits the noise $\textr{\vec{\varepsilon}}$ corrupting $\vec{y}$, and so this data-consistency procedure is best used in the low-noise regime.
\textb{In the presence of significant noise, the dual-decomposition approach \cite{Chen:ECCV:20} may be more appropriate.}

In the accelerated MRI formulation \eqref{kc}-\eqref{A}, the matrix $\vec{A}$ is itself an orthogonal projection matrix, so that, in \eqref{xhat},
\begin{eqnarray}
\vec{I}-\vec{A}^+\vec{A}
&=& \blkdiag\!\big\{ \vec{F}\herm\tvec{P}\tran\tvec{P}\vec{F},\dots,
    \vec{F}\herm\tvec{P}\tran\tvec{P}\vec{F} \big\} 
\label{eq:IAA},\qquad 
\end{eqnarray} 
where $\tvec{P}\in\Real^{(D-M)\times D}$ is the sampling matrix for the non-measured k-space.
Also, $\vec{y}$ is in the row-space of $\vec{A}$, so
\begin{align} 
\vec{A}^+\vec{y} 
&= \vec{y} 
\end{align} 
in \eqref{xhat}.
Figure \ref{fig:architecture} illustrates the overall procedure, using ``data consistency'' to describe \eqref{xhat} and \textb{``nullspace projection'' to describe \eqref{ui}. 
In \secref{ablation}, we quantitatively demonstrate the improvements gained from designing our CNF to estimate only the nullspace component.}

%

\subsection{Architecture} \label{sec:Architecture}
The backbone of $g_{\vec{\theta}}$ is a UNet \cite{Ronneberger:MICCAI:15} that mimics the design in \cite{Zbontar:18}, with 4 pooling layers and 128 output channels in the first convolution layer. 
The first layer was modified to accept complex-valued coil images. 
The inputs have $2C$ channels, where $C$ is the number of coils, each with a real and imaginary component. 
The outputs of the final feature layer of the UNet are processed by a feature-extraction network with $L$ convolution layers. 
Together, the feature extraction network and the UNet make up our conditioning network $g_{\vec{\theta}}$. 
The output of each convolution layer is fed to conditional coupling blocks of the corresponding layer in $h_{\vec{\theta}}$. 

For the flow model $h_{\vec{\theta}}$, we adopt the multi-scale RealNVP \cite{Dinh:ICLR:17} architecture. 
This construction utilizes $L$-layers and $B$-flow steps in each layer. 
A flow step consists of an activation normalization \cite{Kingma:NIPS:18}, a fixed $1 \times 1$ orthogonal convolution \cite{Ardizzone:19}, and a conditional coupling block \cite{Ardizzone:21}. 
Each layer begins with a checkerboard downsampling (squeeze layer) \cite{Dinh:ICLR:17} and a transition step made up of an activation normalization and $1 \times 1$ convolution. 
Layers end with a split operation that sends half of the channels directly to the output on the latent side. 
For all experiments, we use $L=3$ and $B=20$. 
The full architecture of $h_{\vec{\theta}}$ is specified in Fig.\ \ref{fig:architecture}.

\textb{Although the code that accompanies \cite{Denker:JI:21} gives a built-in mechanism to scale their flow architecture to accommodate an increased number of input and output channels, we find that this mechanism does not work well (see \secref{ablation}). 
Thus, in addition to incorporating nullspace learning, we redesign several aspects of the flow architecture and training.
First, to prevent the number of flow parameters from growing unreasonably large, our flow uses fewer downsampling layers (3 vs 6) but more flow steps per downsampling layer (20 vs 5), and we utilize one-sided (instead of two-sided) affine coupling layers.
Second, to connect the conditioning network to the flow, Denker et al.\ \yrcite{Denker:JI:21} used a separate CNN for each flow layer and adjusted its depth to match the flow-layer dimension. 
We use a single, larger CNN and feed its intermediate features to the flow layers with matched dimensions, further preventing an explosion in the number of parameters. 
Third, our conditioning network uses a large, pretrained UNet, whereas Denker et al.\ \yrcite{Denker:JI:21} used a smaller untrained UNet.
With our modifications, we grow the conditional network more than the flow network, which allows the CNF to better handle the high dimensionality of complex-valued, multi-coil data.}

\subsection{Data} \label{sec:Data}

We apply our network to two datasets: the fastMRI knee and fastMRI brain datasets \cite{Zbontar:18}. 
For the knee data, we use the non-fat-suppressed subset, giving 17286 training and 3592 validation images. 
We compress the measurements to $C=8$ complex-valued virtual coils using \cite{Zhang:MRM:13} and crop the images to $320 \times 320$ pixels. 
The sampling mask is generated using the golden ratio offset (GRO) \cite{Joshi:22} Cartesian sampling scheme with an acceleration rate $R=4$ and autocalibration signal (ACS) region of 13 pixels. 
We create the ZF coil-image vectors $\vec{y}$ by applying the mask and inverse Fourier transform to the fully sampled $\vec{k}_c$ given by the fastMRI dataset to obtain $\vec{y}_c=\vec{F}\herm\vec{P}\tran\vec{Pk}_c$ for all $c$, and then stack the coils to obtain $\vec{y}=[\vec{y}_1\tran,\dots,\vec{y}_C\tran]\tran$. 
We create the ground-truth coil-image vectors $\vec{x}\true$ using the same procedure but without the mask, i.e., $\vec{x}_c=\vec{F}\herm\vec{k}_c$ and $\vec{x}\true=[\vec{x}_1\tran,\dots,\vec{x}_C\tran]\tran$.

With the brain data, we use the T2-weighted images and take the first 8 slices of all volumes with at least 8 coils. 
This provides 12224 training and 3352 validation images. 
The data is compressed to $C=8$ virtual coils \cite{Zhang:MRM:13} and cropped to $384 \times 384$ pixels. 
The GRO sampling scheme is again used with an acceleration rate $R=4$ and a 32-wide ACS region. For both methods, the coil-sensitivity maps are estimated from the ACS region using ESPIRiT \cite{Uecker:MRM:14}. 
All inputs to the network are normalized by the 95th percentile of the ZF magnitude images.

\subsection{Training} \label{sec:Training}
For both \textr{datasets}, we first train \textr{the UNet in $g_{\vec{\theta}}$} with an additional $1 \times 1$ convolution layer to get the desired $2C$ channels. 
We train \textr{the UNet} to minimize the mean-squared error (MSE) from the \textb{nullspace projected} targets \textb{$\{\vec{u}\of{i}\}_{i=1}^{N}$} for 50 epochs with batch size 8 and learning rate 0.003. 
Then, we remove the final $1 \times 1$ convolution and jointly train $g_{\vec{\theta}}$ and $h_{\vec{\theta}}$ for 100 epochs to minimize the negative log-likelihood (NLL) loss of the \textb{nullspace projected} targets.
For the brain data, we use batch size 8 and learning rate 0.0003. 
For the knee data, we use batch size 16 with learning rate 0.0005. 
All experiments use the Adam optimizer \cite{Kingma:ICLR:15} with default parameters $\beta_{1} = 0.9$ and $\beta_{2} = 0.999$. 
\textr{The full training takes about 4 days on 4 Nvidia V100 GPUs.}

\subsection{Comparison Methods}
We compare against other methods that are capable of generating posterior samples for accelerated MRI. 
For the fastMRI brain data, we present results for the CGAN from \cite{Adler:18} and the Langevin method from \cite{Jalal:NIPS:21}.
For the fastMRI knee data, we present results for the ``Score'' method from \cite{Chung:MIA:22} and the ``sCNF'' method from \cite{Denker:JI:21}.

For the CGAN, we utilize a UNet-based generator with 4 pooling layers and 128 output channels in the initial layer and a 5-layer CNN network for the discriminator. 
The generator takes $\vec{y}$ concatenated with a latent vector $\vec{z}$ as input. 
The model is trained with the default loss and hyperparameters from \cite{Adler:18} for 100 epochs with a learning rate of 0.001. 
For the Langevin method, we use the authors' implementation but with the GRO sampling mask described in \secref{Data}.

The Score method is different than the other methods in that it assumes that the k-space measurements $\vec{k}_c$ are constructed from true coil images $\vec{x}\true$ with magnitudes affinely normalized to the interval $[0,1]$ and phases normalized to $[0,1]$ radians. \textr{Although this normalization cannot be enforced on prospectively undersampled MRI data, Score fails without this normalization.}
So, to evaluate Score, we normalize each $\vec{k}_c$ using knowledge of the ground-truth $\vec{x}_c$, run Score, and un-normalized its output $\hvec{x}_c$ for comparison with the other methods.
Since the Score paper \cite{Chung:MIA:22} used RSS combining to compute $\hvec{i}$, we do the same.
For the Score method, we use $T\!=\!200$ iterations and not the default value of $T\!=\!2000$.
This is because, when using posterior-sample averaging (see \secref{eval}), the PSNR computed using $200$ iterations is better than with $2000$. 

The sCNF method works only on single-coil magnitude data, and so we convert our multi-coil data to that domain in order to evaluate sCNF.
\textb{To do this, we apply RSS \eqref{rss} to ZF coil-images $\vec{y}$ and repeat the process for the true coil images $\vec{x}\true$.} 
Using those magnitude images, we train sCNF for 300 epochs with learning rate 0.0005 and batch size 32. 

\begin{figure}[t]
\includegraphics[width=1.0\linewidth,trim=7 7 0 7,clip]{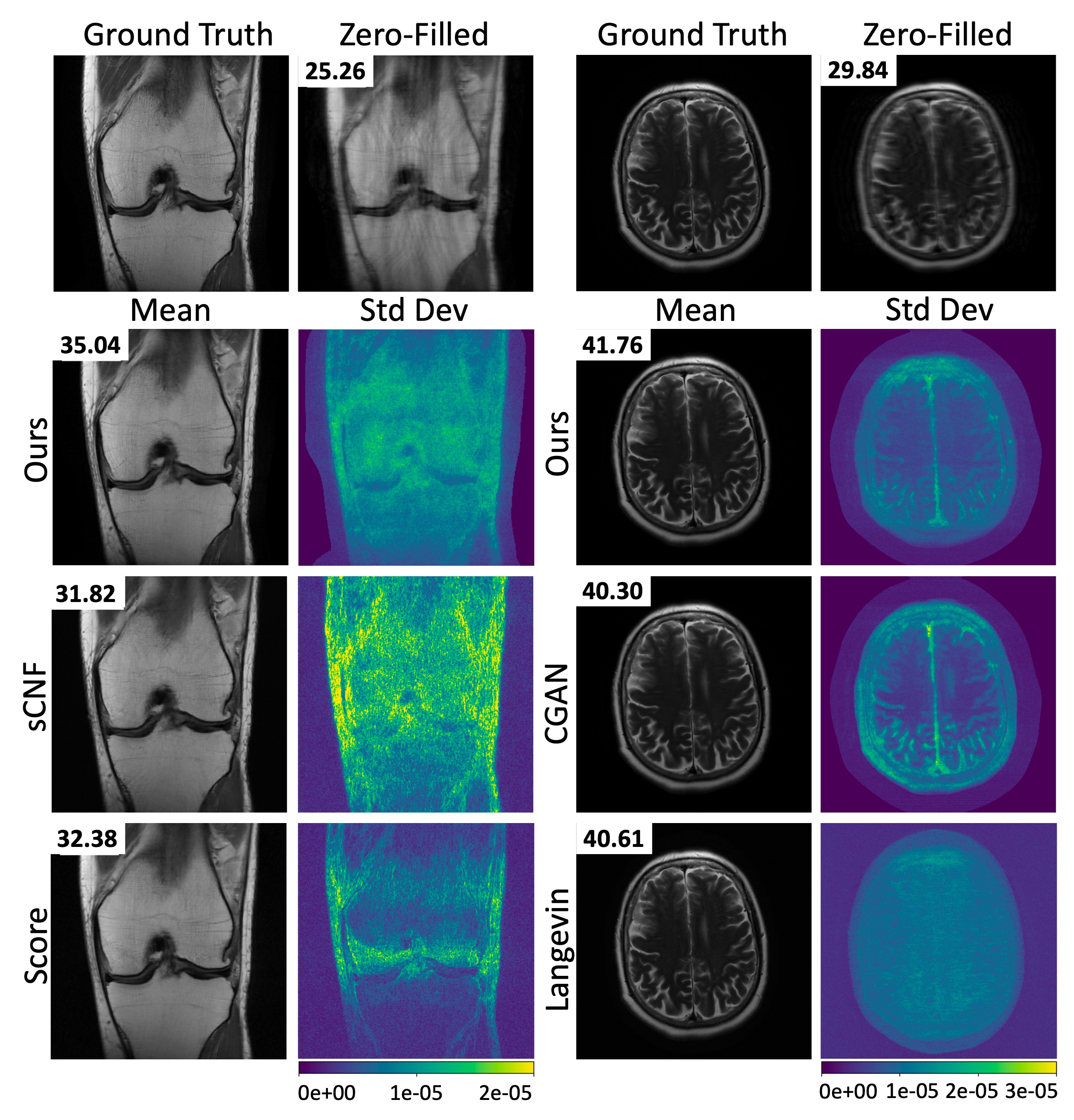}
\vspace{-7mm}
\caption{Mean images and pixel-wise standard-deviation maps computed from 8 and 32 posterior samples for the brain images and knee datasets, respectively. The standard-deviation maps show which pixels have the greatest reconstruction uncertainty. The corresponding PSNR is shown on each reconstruction.}
\label{fig:mean_imgs}
\end{figure}

\subsection{Evaluation} \label{sec:eval}
We report results for several different metrics, including peak-signal-to-noise ratio (PSNR), structural-similarity index (SSIM) \cite{Wang:04:SSIM}, Fr\'echet Inception Score (FID) \cite{Heusel:NIPS:17}, and conditional FID (cFID) \cite{Soloveitchik:21}.
PSNR and SSIM were computed on the average of $P$ posterior samples $\{\vec{i}_p\}_{p=1}^P$, i.e., 
\begin{align}
\hvec{i}_{(P)} \defn \frac{1}{P} \sum_{p=1}^{P} \hvec{i}_{p}
\label{eq:hatiP}
\end{align}
to approximate the posterior mean, while FID and cFID were evaluated on individual posterior samples $\hvec{i}_p$. 
By default, we compute all metrics using magnitude reconstructions $|\hvec{i}|$ rather than the complex-valued reconstructions $\hvec{i}$, in part because competitors like sCNF generate only magnitude reconstructions, but also because this is typical in the MRI literature (e.g., the fastMRI competition \cite{Zbontar:18}).
So, for example, PSNR is computed as 
\begin{equation}
\text{PSNR}
\defn 10 \log_{10} \bigg( \frac{D\max_d |[\vec{i}\true]_d|^{2}}{\big\| |\vec{\hat{i}}_{(P)}| - |\vec{i}\true| \big\|_2^2} \bigg)
\label{eq:PSNR} ,
\end{equation}
where $[\cdot]_d$ extracts the $d$th pixel.
For FID and cFID, we use the embeddings of VGG-16 \cite{Simonyan:14} as \cite{Kastryulin:22} found that this helped the metrics better correlate with the rankings of radiologists. 

For the brain data, we compute all metrics on 72 random test images in order to limit the Langevin \textr{image generation} time to 4 days. 
We generate complex-valued images using the coil-combining method in \eqref{sense} and use $P=32$ posterior samples to calculate cFID\textsuperscript{1}, FID\textsuperscript{1}, PSNR, and SSIM. 
(For the reference statistics of FID, we use the entire training dataset.)
Because FID and cFID are biased by small sample sizes, we also compute FID\textsuperscript{2} and cFID\textsuperscript{2} with 2484 test samples and $P=8$ for our method and the CGAN. 

With the knee data, we follow a similar evaluation procedure except that,
to comply with the evaluation steps of Score, we generate magnitude-only signals using the root-sum-of-square (RSS) combining from \eqref{rss}.
Also, we computed metrics on 72 randomly selected slices in order to bound the \textr{image generation} time of Score to 6 days with $P=8$. 
We use $P=8$ for all metrics, but for FID\textsuperscript{2} and cFID\textsuperscript{2}, we use 2188 test samples. 

When computing \textr{inference} time for all methods, we use a single Nvidia V100 with 32GB of memory and evaluate the time required to generate one posterior sample.

\section{Results}

\begin{figure*}[t!]
\centerline{\includegraphics[width=0.99\linewidth]{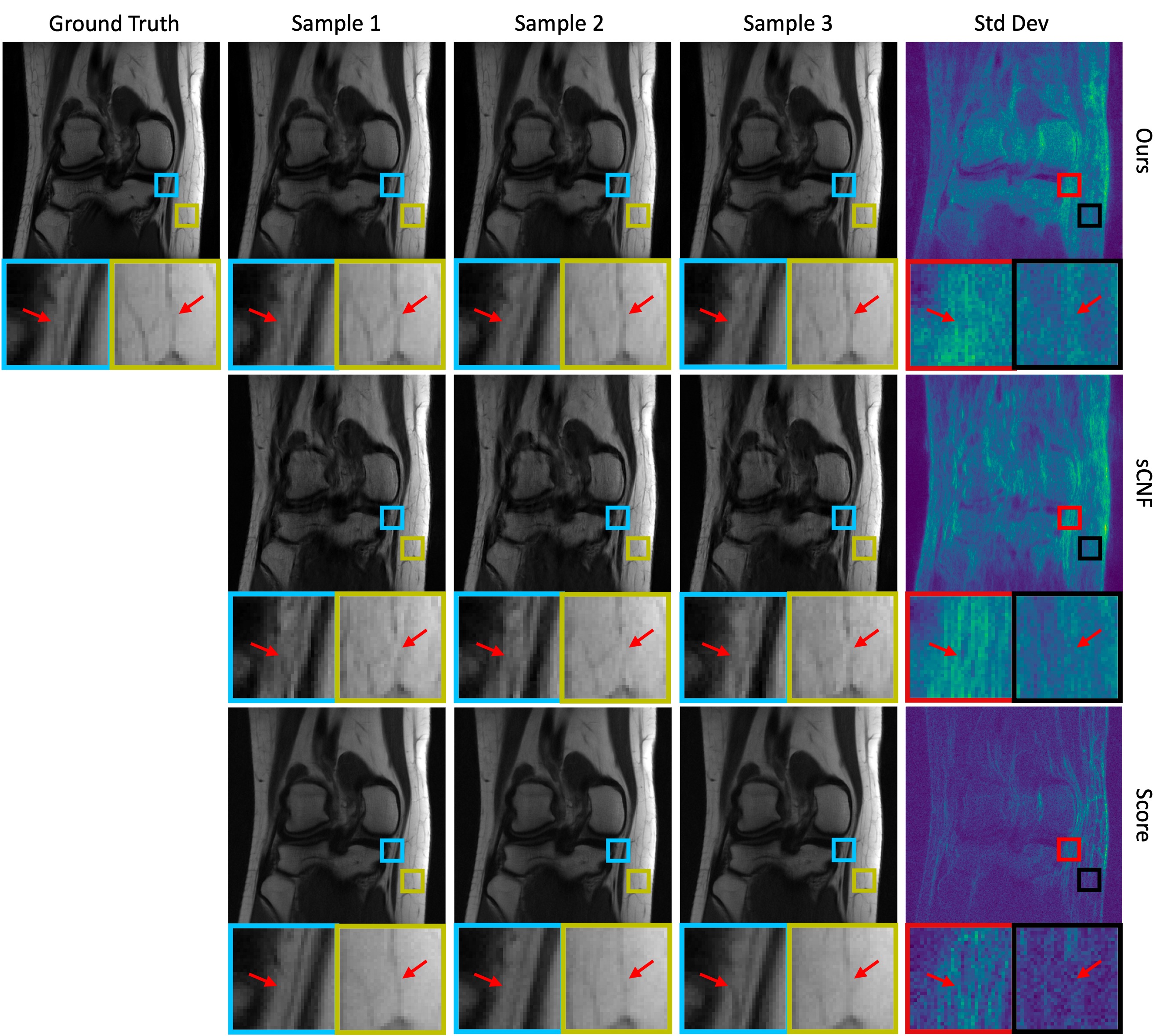}}
\vspace{\spc}
\caption{Examples of posterior samples and standard-deviation maps for the knee data. The samples show important structural variations. This demonstrates the advantages of generating multiple reconstructions and computing a pixel-wise standard-deviation map.}
\label{fig:posteriors}
\end{figure*}

\begin{table*}[t]
\centering
\resizebox{1.4\columnwidth}{!}{%
\begin{tabular}{|l|l|l|l|l|l|l|l|}
\hline
   Model & PSNR (dB) $\!\uparrow$  & SSIM $\!\uparrow$ & FID\textsuperscript{1} $\!\downarrow$ & FID\textsuperscript{2} $\!\downarrow$ & cFID\textsuperscript{1} $\!\downarrow$ & cFID\textsuperscript{2} $\!\downarrow$ & Time \\ \hline \hline
Score & 34.15 $\pm$ 0.19   & 0.8764 $\pm$ 0.0036 & \bf 4.49 & --- & 4.49 & --- & 15  min\\ \hline
sCNF & 32.93 $\pm$ 0.17   &  0.8494 $\pm$ 0.0047 & 7.32 & 5.78 & 8.49 & 6.51 & \bf 66 ms\\ \hline 
Ours & \bf 35.23 $\pm$ 0.22   & \bf 0.8888 $\pm$ 0.0046 & 4.68 & \bf 2.55 & \bf 3.96 & \bf 2.44 & 108  ms \\ \hline
\end{tabular}}
\vspace{-1mm}
\caption{Average performance on non-fat-suppressed fastMRI knee data, with standard error reported after the $\pm$. PSNR, SSIM, FID\textsuperscript{1}, and cFID\textsuperscript{1} are computed for 72 test images and $P=8$ posterior samples. FID\textsuperscript{2}, and cFID\textsuperscript{2} are computed for 2188 test samples and $P=8$ posterior samples. Time to the generation of one posterior sample.}
\label{tab:knee_metrics}
\end{table*}

\begin{table*}[t]
\centering
\resizebox{1.45\columnwidth}{!}{%
\begin{tabular}{|l|l|l|l|l|l|l|l|}
\hline
   Model & PSNR (dB) $\!\uparrow$  & SSIM $\!\uparrow$ & FID\textsuperscript{1} $\!\downarrow$ & FID\textsuperscript{2} $\!\downarrow$ & cFID\textsuperscript{1} $\!\downarrow$ & cFID\textsuperscript{2} $\!\downarrow$ & Time \\ \hline \hline
Langevin  & 37.88 $\pm$ 0.41   & 0.9042 $\pm$ 0.0062 & 6.12 & --- & 5.29 & --- & 14 min\\ \hline 
CGAN  & 37.28 $\pm$ 0.19   & 0.9413 $\pm$ 0.0031 & 5.38 & 4.06 & 6.41 & 4.28 & \bf 112 ms\\ \hline
Ours  & \bf 38.85 $\pm$ 0.23   & \bf 0.9495 $\pm$ 0.0012 & \bf 4.13 & \bf 2.37 & \bf 4.15 & \bf 2.44 & 177  ms\\ \hline
\end{tabular}}
\vspace{-1mm}
\caption{Average performance on non-fat-suppressed fastMRI brain data, with standard error reported after the $\pm$. PSNR, SSIM, FID\textsuperscript{1}, and cFID\textsuperscript{1} are computed for 72 test images and $P=32$ posterior samples. FID\textsuperscript{2} and cFID\textsuperscript{2} are computed using 2484 test samples and $P=8$. Time to the generation of one posterior sample.}
\label{tab:brain_metrics}
\end{table*}

\tabref{knee_metrics} reports the quantitative metrics for the knee dataset. 
It shows that our method outperforms sCNF by a significant margin in all metrics except inference time. 
By using information from multiple coils and a more advanced architecture, our method shows the true competitive potential of CNFs in realistic accelerated MR imaging. 

\tabref{knee_metrics} also shows that our method surpasses Score in all metrics except FID\textsuperscript{1}, even though Score benefited from impractical ground-truth normalization. 
Compared to Score, our method generated posterior samples $8000 \times$ faster.  
Furthermore, our method (and sCNF) will see a speedup when multiple samples are generated because the conditioning network $g_{\vec{\theta}}$ needs to be evaluated only once per $P$ generated samples for a given $\vec{y}$.
For example, with the knee data, we are able to generate $P=32$ samples in $1.41$ seconds, corresponding to $44$ milliseconds per sample, which is a $2.5\times$ speedup over the value reported in \tabref{knee_metrics}.

\tabref{brain_metrics} reports the quantitative results for the brain dataset. 
The table shows that we outperform the Langevin and CGAN methods in all benchmarks except inference time. 
While our method is a bit slower than the CGAN, it is orders of magnitude faster than the Langevin approach.


We show the mean images and standard-deviation maps for the fastMRI knee and brain experiments in Fig.\ \ref{fig:mean_imgs}. 
For the knee data, our method captures texture more accurately than the sCNF method and provides a sharper representation than the Score method. 
All of the brain methods provide a visually accurate representation to the ground truth, but the Langevin method provides a \textb{more diffuse} variance map, with energy spread throughout the image. 

In Fig.\ \ref{fig:posteriors}, we plot multiple posterior samples, along with zoomed-in regions, to illustrate the changes across independently drawn samples for each method. 
The standard-deviation maps are generated using $P=8$ posterior samples, three of which are shown. 
From the zoomed-in regions, it can be seen that several samples are consistent with the ground truth while others are not (although they may be consistent with the measured data). 
Regions of high posterior variation can be flagged from visual inspection of the standard-deviation map and further investigated through viewing multiple posterior samples for improved clinical diagnoses.

Our method presents observable, realistic variations of small anatomical features in the zoomed-in regions. 
The variations are also registered in the standard-deviation map. 
Both the posterior samples and the standard-deviation map could be used by clinicians to assess their findings. 
Comparatively, our method demonstrates variation that is spread across the entire image, while in the Score method, the variation is mostly localized to small regions. 
\textb{Since it is difficult to say which standard-deviation map is more useful or correct, the interpretation of these maps could be an interesting direction for future work.}
The sCNF also demonstrates variation, but it is mostly driven by residual aliasing artifacts.




\begin{figure}[t]
\centerline{\includegraphics[width=1.0\linewidth,trim=10 10 10 10,clip]{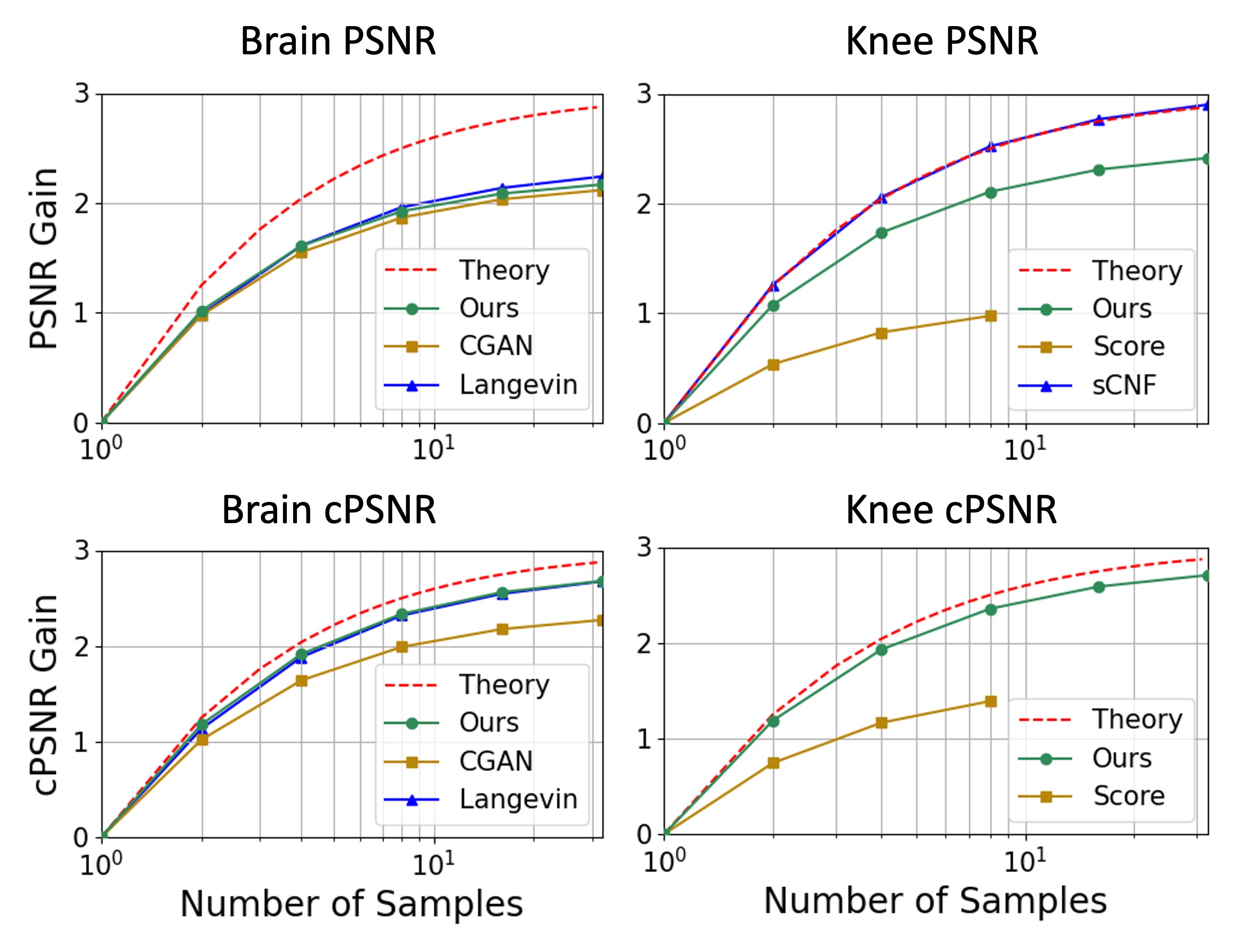}}
\vspace{\spc}
\caption{The gain in (magnitude) PSNR and complex PSNR of the $P$-sample mean estimate $\hvec{i}_{(P)}$ versus $P$, for both brain and knee data. Note the $\approx$ 3 dB increase as $P$ grows from 1 to infinity.}
\label{fig:psnr}
\end{figure}


\subsection{PSNR Gain versus $P$}

It is well known that the minimum MSE (MMSE) estimate of $\vec{i}$ from $\vec{y}$ equals the conditional mean $\E\{\vec{i}|\vec{y}\}$, i.e., the mean of the posterior distribution $p_{i|y}(\cdot|\vec{y})$.
Thus, one way to approximate the MMSE estimate is to generate many samples from the posterior distribution and average them, as in \eqref{hatiP}.
Bendel et al.\ \yrcite{Bendel:22} showed that the MSE
\begin{equation}
\mathcal{E}_{P} \defn \E\big[ \| \hvec{i}_{(P)} - \vec{i}\true \| ^{2}_{2} \big| \vec{y} \big]
\label{eq:EP}
\end{equation}
of the $P$-posterior-sample average $\hvec{i}_{(P)}$ obeys 
$\mathcal{E}_{1}/\mathcal{E}_{P} = 2P/(P+1)$.
So, for example, the SNR increases by a factor of two as $P$ grows from 1 to $\infty$.  
The same thing should happen for PSNR, as long as the PSNR definition is consistent with \eqref{EP}.
For positive signals (i.e., magnitude images) the PSNR definition from \eqref{PSNR} is consistent with \eqref{EP}, but for complex signals we must use ``complex PSNR'' 
\begin{equation}
\text{cPSNR}
\defn 10 \log_{10} \bigg( \frac{D\max_d |[\vec{i}\true]_d|^{2}}{\| \vec{\hat{i}}_{(P)} - \vec{i}\true \|_2^2} \bigg)
\label{eq:cPSNR} .
\end{equation}
As RSS combining provides only a magnitude estimate, we compute the coil-combined estimate for our method and Score to evaluate cPSNR behavior for the knee dataset.

One may then wonder whether a given approximate posterior sampler has a PSNR gain versus $P$ that matches the theory.
In Fig.\ \ref{fig:psnr}, we answer this question by plotting the PSNR gain and the cPSNR gain versus $P\in\{1,2,4,8,16,32\}$ for the various methods under test 
(averaged over all 72 test samples). 
There we see that our method's cPSNR curve matches the theoretical curve well for both brain and knee data.
As expected, our (magnitude) PSNR curve does not match the theoretical curve. 
The cPSNR curves of the Score and CGAN methods fall short of the theoretical curve by a large margin, but interestingly, the Langevin method's cPSNR curve matches ours almost perfectly. 
sCNF's PSNR gain curve matches the theoretical one almost perfectly, which provides further empirical evidence that CNF methods accurately sample from the posterior distribution. 

%

\subsection{Ablation Study} \label{sec:ablation}

\begin{table}[t!]
\centering
\resizebox{\columnwidth}{!}{%
\begin{tabular}{|l|l|l|l|l|}
\hline
  Model & PSNR (dB) $\!\uparrow$  & SSIM $\!\uparrow$ & FID\textsuperscript{2} $\!\downarrow$ & cFID\textsuperscript{2} $\!\downarrow$\\ \hline \hline
\cite{Denker:JI:21} & 17.61 $\pm$ 0.20   & 0.6665 $\pm$ 0.0072 & 16.02 &  16.68 \\ \hline
+ Data Consistency & 27.27 $\pm$ 0.21   & 0.7447 $\pm$ 0.0061 & 16.92 &  18.56 \\ \hline
+ Architectural Changes & 33.87 $\pm$ 0.23   & 0.8715 $\pm$ 0.0049 & 4.48 &  4.50 \\ \hline
+ Nullspace Learning & \bf 35.23 $\pm$ 0.22   & \bf 0.8888 $\pm$ 0.0046 &  \bf 2.55 &  \bf 2.44  \\ \hline
\end{tabular}}
\vspace{-2mm}
\caption{Ablation Study: Performance on non-fat-suppressed fastMRI knee data, with standard error reported after the $\pm$. Each line adds a new contribution to the model of the previous line. Metrics are computed as described in \secref{eval}}
\label{tab:ablation}
\end{table}

To evaluate the impact of our contributions to CNF architecture and training design, we perform an ablation study using the fastMRI knee dataset. 
We start with the baseline model in \cite{Denker:JI:21}, modified to take in 16 channels instead of 1, and scale it up using the built-in mechanism in the author's code. 
We train this model for 300 epochs with batch size 32 and learning rate 0.0001 to minimize the NLL of the multicoil targets $\{\vec{x}\of{i}\}$, since higher learning rates were numerically unstable. 
\tabref{ablation} shows what happens when we add each of our contributions. 
First, we add data consistency \eqref{xhat} to the evaluation of the baseline.
We then add the architectural changes described in \secref{Architecture}, 
and finally we add nullspace learning to arrive at our proposed method. 
From \tabref{ablation}, it can be seen that each of our design contributions
yielded a significant boost in performance, and that nullspace learning was a critical ingredient in our outperforming the Score method in \tabref{knee_metrics}.
For this ablation study, all models were trained following the procedure outlined in \secref{Training} (except for the learning rate of the baseline).

\subsection{Maximum a Posteriori (MAP) Estimation}

\begin{figure}[t!]
\centerline{\includegraphics[width=\linewidth,trim=0 10 0 10,clip]{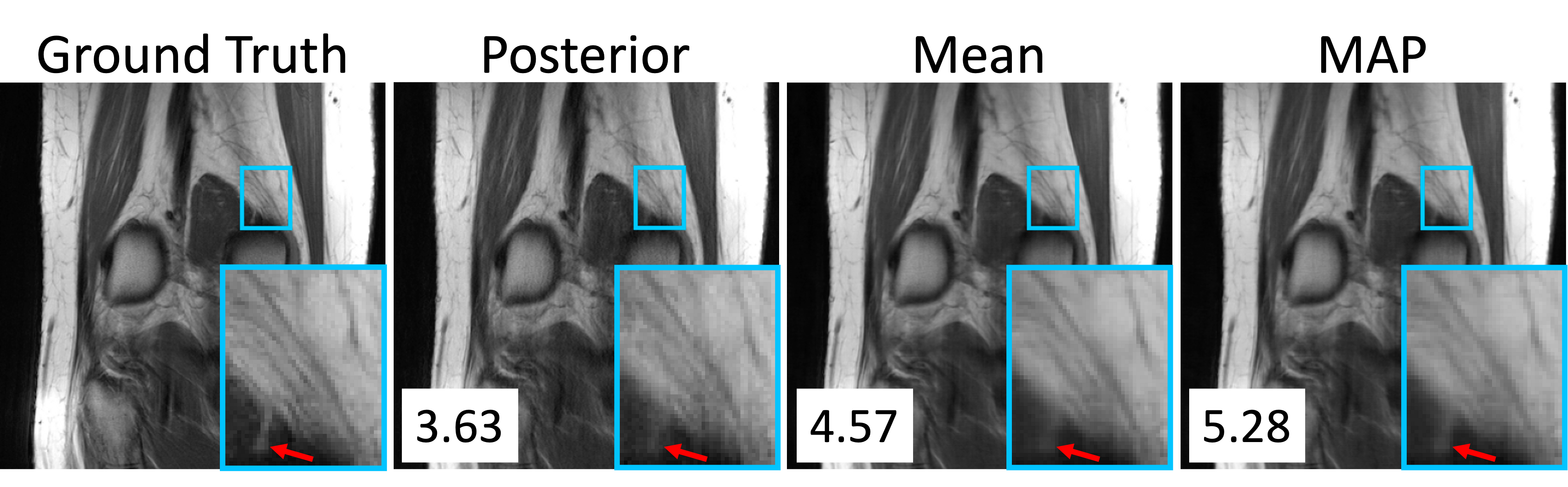}}
\vspace{\spc}
    \caption{Examples of a ground-truth image, one posterior sample, an average of $P=8$ posterior samples, and a MAP estimate. The log posterior density in units of bits-per-dimension is shown in the bottom right corner of each image.}
\label{fig:map}
\end{figure}

Because CNFs can evaluate the posterior density of a signal hypothesis (recall \eqref{px|y}), they can be used for posteriori (MAP) estimation, unlike CGANs. 

Due to our data-consistency step \eqref{xhat}, we find the MAP estimate of $\vec{x}$ using
\begin{align}
\hvec{x}\map
&= \hvec{u}\map + \vec{A}^+\vec{y} \\
\hvec{u}\map
&= \arg\max_{\vec{u}\in\nullsp(\vec{A})} \ln \hat{p}_{u|y}(\vec{u}|\vec{y}) 
\label{eq:umap} .
\end{align}
Note the CNF output $\vec{u}$ is constrained to the nullspace of $\vec{A}$.
From \eqref{IAA}, this nullspace is spanned by the columns of 
\begin{align}
\vec{W} 
\defn \blkdiag\big\{\vec{F}\herm\tvec{P}\tran, \dots, \vec{F}\herm\tvec{P}\tran \big\}
\label{eq:W},
\end{align}
which are orthonormal, and so $\hvec{u}\map=\vec{W}\tvec{k}\map$ with
\begin{align}
\tvec{k}\map 
&= \arg\max_{\tvec{k}} \ln \hat{p}_{u|y}(\vec{W}\tvec{k} | \vec{y};\vec{\theta}) \\
&= \arg\max_{\tvec{k}} \Bigg[ \ln p_{z}(\bar{h}_{\vec{\theta}}^{-1}(\vec{W}\tvec{k},\vec{y})) 
\nonumber\\&\quad\mbox{}
+\ln \!\bigg| \!\det \!\bigg( \frac{\partial \bar{h}_{\vec{\theta}}^{-1}(\tvec{u},\vec{y})}{\partial \tvec{u}}\bigg|_{\tvec{u}=\vec{W}\tvec{k}}  \bigg) \bigg| \Bigg] .
\end{align}
For this maximization, we use the Adam optimizer with 5000 iterations and a learning rate of $1\times 10^{-8}$. 
Above, $\tvec{k}$ can be recognized as the unmeasured k-space samples. 

In Figure \ref{fig:map}, we show an example of a MAP estimate along with the ground truth image, one sample from the posterior, a $P=8$ posterior-sample average, and their corresponding log-posterior-density values. 
As expected, the MAP estimate has a higher log-posterior-density that the other estimates. 
Visually, the MAP estimate is slightly sharper than the sample average but contains less texture details than the single posterior sample.

\section{Conclusion}
In this work, we present the first conditional normalizing flow for posterior sample generation in multi-coil accelerated MRI.
To do this, we designed a novel conditional normalizing flow (CNF) that infers the signal component in the measurement operator's nullspace, whose outputs are later combined with information from the measured space.
In experiments with fastMRI brain and knee data, we demonstrate improvements over existing posterior samplers for MRI.
Compared to score/Langevin-based approaches, our inference time is four orders-of-magnitude faster.
We also illustrate how the posterior samples can be used to quantify uncertainty in MR imaging. 
This provides radiologists with additional tools to enhance the robustness of clinical diagnoses. 
We hope this work motivates additional exploration of posterior sampling for accelerated MRI.

\section*{Acknowledgements}

This work was supported in part by the National Institutes of Health under Grant R01-EB029957.

\clearpage
\bibliography{macros_abbrev,machine,mri,misc,sparse}
\bibliographystyle{icml2023}

\newpage
\appendix
\onecolumn

\section{Accelerated MRI Simulation Procedure}
We outline the procedure for simulating the accelerated MRI problem. 
The fastMRI datasets provide the fully sampled multi-coil k-space, i.e., $\{\vec{k}_c\}_{c=1}^C$ with $M=D$.  
To obtain the ground truth coil-images $\{\vec{x}_c\}_{c=1}^C$, we take the inverse Fourier transform of the fully sampled k-space measurement, i.e., $\vec{x}_c=\vec{F}\herm\vec{k}_c$, wherein we assume that the noise $\vec{\epsilon}_c$ in \eqref{kc} is negligible. 
To obtain the zero-filled images $\vec{y}_c$, we take the inverse Fourier transform after masking the fully-sampled k-space measurement $\vec{k}_c$, i.e., $\vec{y}_c=\vec{F}\herm\vec{P}\tran\vec{Pk}_c$. 
This procedure is illustrated in Fig. \ref{fig:mri_sim}. 
In real-world accelerated MRI, the data acquisition process would collect masked k-space $\vec{Pk}_c$ directly.

\begin{figure}[h!]
\centerline{\includegraphics[width=\linewidth]{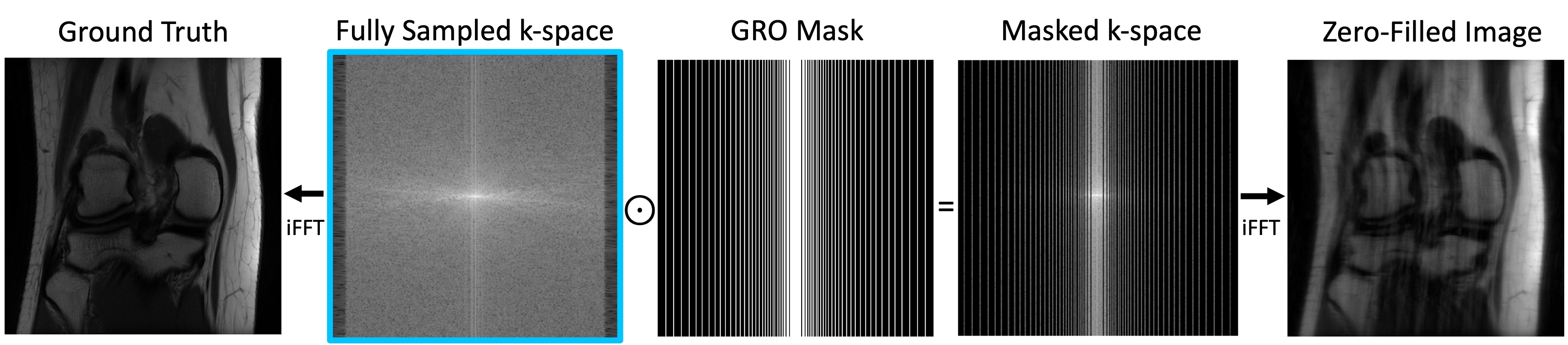}}
    \caption{A visual illustration of simulating accelerated MRI. Given the fully sampled k-space $\vec{k}_c$ \textb{highlighted in blue}, we obtain the ground truth $\vec{x}_c$ by applying the inverse Fourier transform $\vec{F}\herm$. The zero-filled image $\vec{y}_c$ is acquired by applying the sampling mask $\vec{P}\tran\vec{P}$ to fully sampled $\vec{k}_c$ and then taking the inverse Fourier Transform $\vec{F}\herm$.}
\label{fig:mri_sim}
\end{figure}

\section{Implementation Details}
For our machine learning framework, we use PyTorch \cite{pytorch} and PyTorch lightning \cite{lightning}. 
To implement the components of the CNF, we use the Framework for Easily Invertible Architectures (FrEIA) \cite{Ardizzone:github:18}. 
For the Score, sCNF, and Langevin methods, we utilize the authors' implementations at \cite{Chung:github:22}, \cite{Denker:github:21}, and \cite{Jalal:github:21}, respectively. 
ESPIRiT coil-estimation and coil-combining are implemented using the SigPy package \cite{Ong:SigPy:19}.

\clearpage
\section{Brain Posterior Samples}
\begin{figure}[h!]
\centerline{\includegraphics[width=0.99\linewidth]{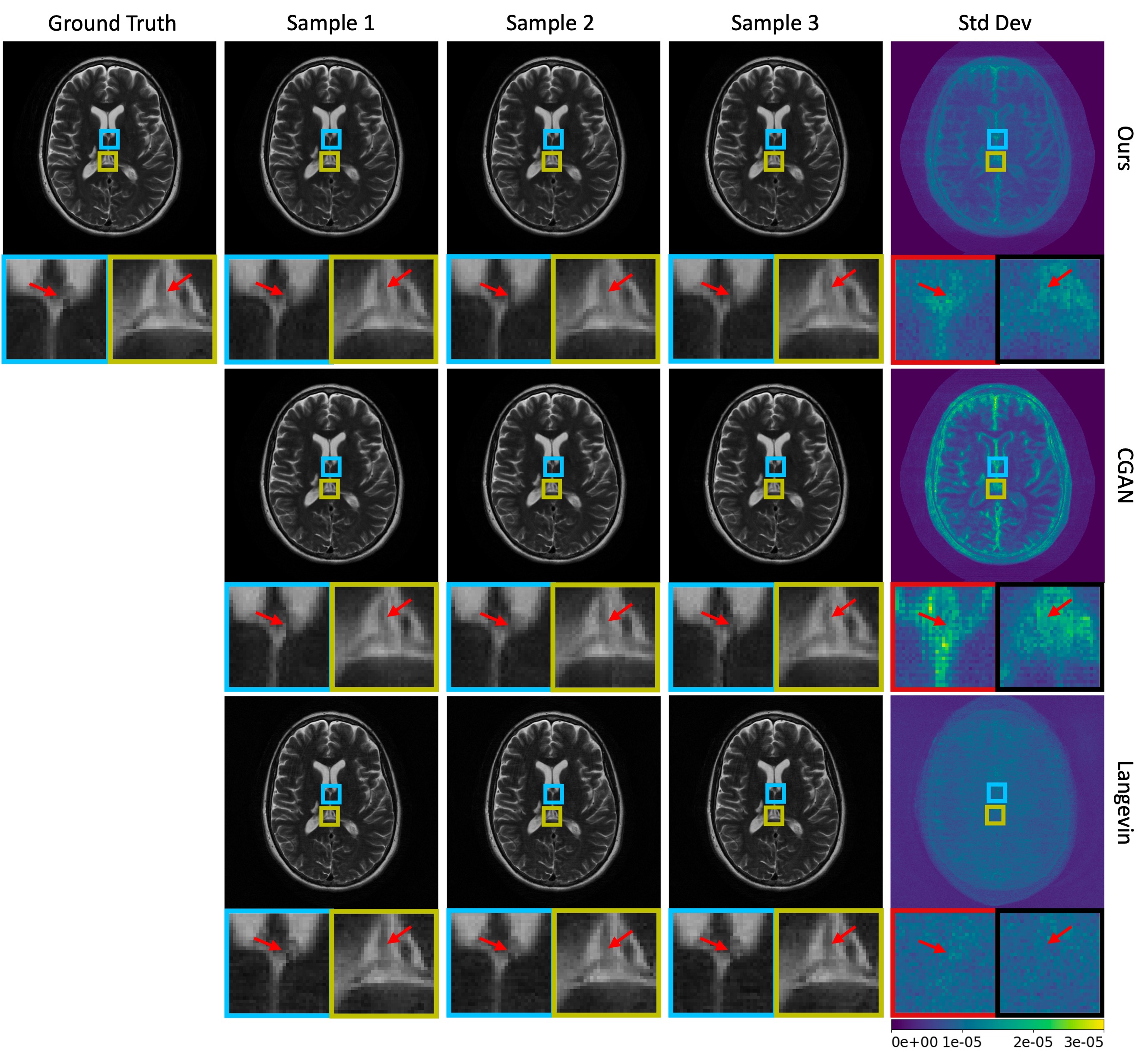}}
    \caption{Examples of posterior samples and standard-deviation maps for the brain images, both with zoomed regions.}
\label{fig:posteriors2}
\end{figure}

\end{document}